\documentclass[twocolumn,showpacs,amsfonts,amsmath,amssymb,aps,pra,floatfix]{revtex4}
\usepackage{graphicx}
\usepackage{bm}
\begin{document}
\title{Gravity-induced resonances in a rotating trap}
\author{Iwo Bialynicki-Birula}
\email{birula@cft.edu.pl}
\author{Tomasz Sowi\'nski}
\affiliation{Center for Theoretical Physics, Polish Academy of Sciences\\Al. Lotnik\'ow 32/46, 02-668 Warsaw, Poland}

\begin{abstract} 
It is shown that in an anisotropic harmonic trap that rotates with the properly chosen rotation rate, the force of gravity leads to a resonant behavior. Full analysis of the dynamics in an anisotropic, rotating trap in 3D is presented and several regions of stability are identified. On resonance, the oscillation amplitude of a single particle, or of the center of mass of a many-particle system (for example, BEC), grows linearly with time and all particles are expelled from the trap. The resonances can only occur when the rotation axis is tilted away from the vertical position. The positions of the resonances (there are always two of them) do not depend on the mass but only on the characteristic frequencies of the trap and on the direction of the angular velocity of rotation.
\end{abstract}
\pacs{03.75.Kk, 45.30.+s, 45.50.-j}
\maketitle
\section{Introduction}

The main purpose of this work is to expose the role of gravity in the dynamics of particles in a rotating trap. In addition, we present a complete analysis of the stability regions for a rotating trap in 3D. We prove that in the generic case there are three separate regions of stability with different characteristics. Gravity induced resonances are relevant only if they occur in the regions of stability, otherwise, they are swamped by the exponential behavior of trajectories.

Harmonic traps are often used in optics and atomic physics (especially in the form of TOP traps \cite{petrich} in the study of Bose-Einstein condensates) and yet a complete theory of these devices has not been developed. The solution to the problem of a trap rotating around one of the trap axes is effectively two-dimensional and its solution has been known for at least hundred years. In the classic textbook on analytical dynamics by Whittaker \cite{whitt} we find a solution of a mathematically equivalent problem of small oscillations of ``a heavy particle about its position of equilibrium at the lowest point of a surface which is rotating with constant angular velocity about a vertical axis through the point''. The quantum-mechanical counterpart of the Whittaker problem has also been completely solved \cite{linn,oktel} and the statistical mechanics of a classical gas was studied in \cite{gd}.

In this work we present a complete solution to the problem of the motion of a particle moving in a most general anisotropic rotating harmonic trap in 3D and in the presence of gravity. This is an exactly soluble problem but technical difficulties apparently served so far as a deterrent in developing a full description. A full description of the particle dynamics in a rotating anisotropic trap in three dimensions so far has not been given despite a new significance of this problem brought about by experimental and theoretical studies of Bose-Einstein condensates and the accompanying thermal clouds in rotating traps \cite{rzs,mad,sc,gg,ros,cozz,abo,gd}. Explicit formulas describing the complete mode structure in the three-dimensional case are indeed quite cumbersome \cite{cmm} because we deal here with third-order polynomials and on top of that they have rather complicated coefficients. However, many important features may be exhibited without straining the reader's patience. In particular, we can identify various stability regions for an arbitrary orientation of the angular velocity and we can give conditions for a resonance.

The standard arrangement \cite{rzs,mad,sc,gg,ros,cozz,abo,gd} is to choose a vertical axis of rotation of the trap. Slight tiltings of this axis were introduced to excite the scissors modes \cite{gds,marago,smith}. However, for such very small tilting angles the effects described in the present paper would not be noticeable. In the case of a vertical axis of rotation, there are no resonances. The only effect of gravity is a displacement of the equilibrium position. The situation completely changes and new phenomena will occur when the axis of rotation is tilted away from the vertical position. In this generic case, for every anisotropic three-dimensional trap there exist two (not three as one might expect) characteristic frequencies at which resonances occur. The motion in a trap that is rotating at a {\em resonant} frequency will become unbounded and all particles will be expelled from the trap. The position of the resonance does not depend on the mass but only on the characteristic frequencies of the trap and on the direction of the angular velocity.

All our results are valid not only for a single particle but also for the center of mass motion in many-body (classical or quantum) theory since for all quadratic Hamiltonians the center of mass motion completely separates from the internal motion \cite{kohn,dob,cmm}. Therefore, a trap rotating at the resonant frequency will not hold the Bose-Einstein condensate. Owing to the linearity of the equations of motion for a harmonic trap, all conclusions hold both in classical and in quantum theory. A resonant behavior caused by an application of a static force may seem counterintuitive, but it is explained by the fact that in a rotating frame the force of gravity acts as a {\em periodically changing} external force.

\section{Equations of motion}

The best way to analyze the behavior of particles in a uniformly rotating trap is to first perform the transformation to the rotating frame. In this frame the harmonic trap potential is frozen but the force of gravity is rotating with the angular velocity of the trap rotation. In the rotating frame the Hamiltonian has the form
\begin{equation}\label{ham}  {\cal H} = \frac{\bm{p}^{2}}{2m}  + \bm{r}\!\cdot\!\hat{\Omega}\!\cdot\!\bm{p}  + \frac{m}{2}\bm{r}\!\cdot\!\hat{V}\!\cdot\!\bm{r} - m\bm{r}\!\cdot\!\bm{g}(t).
\end{equation}
The potential matrix $\hat{V}$ is symmetric and positive definite. The eigenvalues of this matrix are the squared frequencies of the oscillations in the non-rotating trap. The angular velocity matrix $\hat{\Omega}$ is related to the components of the angular velocity vector through the formula ${\Omega}_{ik}=\epsilon_{ijk}\Omega_j$. The vector of the gravitational acceleration $\bm{g}(t)$, as seen in the rotating frame, can be expressed in the form
\begin{eqnarray}\label{rotg}
{\bm g}(t) = {\bm g}_{\parallel} + {\bm g}_{\perp}\cos(\Omega t) - ({\bm n}\times{\bm g}_{\perp})\sin(\Omega t),
\end{eqnarray}
where ${\bm n}$ denotes the direction and $\Omega$ denotes the length of the angular velocity vector ${\bm\Omega}$. The parallel and the transverse components of the gravitational acceleration vector ${\bm g}={\bm g}(0)$ are defined as ${\bm g}_{\parallel} = {\bm n}({\bm n}\!\cdot\!{\bm g})$ and ${\bm g}_{\perp} = {\bm g}-{\bm n}({\bm n}\!\cdot\!{\bm g})$, respectively. Note that the time-dependent part vanishes when the rotation axis is vertical.

The equations of motion determined by the Hamiltonian (\ref{ham}) have the following form
\begin{subequations}
\begin{eqnarray} \label{3dimeq}
\frac{d{\bm r}(t)}{dt} &=& \frac{{\bm p}(t)}{m} - \hat{\Omega}\!\cdot\!{{\bm r}(t)},\\
\frac{d{\bm p}(t)}{dt} &=& -m\hat{V}\!\cdot\!{{\bm r}(t)} - \hat{\Omega}\!\cdot\!{{\bm p}(t)} + m {\bm g}(t).
\end{eqnarray}
\end{subequations}
These equations describe an oscillator in a rotating frame displaced by a constant force (the longitudinal part of ${\bm g}$) and driven by a periodic force (the transverse part of ${\bm g}$). It is convenient to rewrite the expression (\ref{rotg}) as a real part of a complex function
\begin{eqnarray}\label{rotg1}
 {\bm g}(t) = \Re\left({\bm g}_{\parallel}+({\bm g}_{\perp} +  i({\bm n}\times{\bm g}_{\perp}))e^{i\Omega t}\right).
\end{eqnarray}
In compact notation Eqs.~(\ref{3dimeq}) have the form
\begin{equation} \label{eqnmot}
\frac{d{\cal R}(t)}{dt} = \hat{\cal M}(\Omega)\!\cdot\!{\cal R}(t) + \Re({\cal G}_{\parallel} + {\cal G}_{\perp}{e}^{i\Omega t}),
\end{equation}
where
\begin{eqnarray}
{\cal R}(t) &=& \left(\begin{array}{c} \bm{r}(t) \\ \bm{p}(t) 
\end{array}\right),\;\;
\hat{\cal M}(\Omega) = \left(\begin{array}{cc} - \hat{\Omega} & m^{-1}\hat{I} \\ -m\hat{V} & - \hat{\Omega}\end{array}\right), \\
{\cal G}_{\parallel} &=&  m\left(\begin{array}{c} 0 \\ \bm{g}_{\parallel} \end{array}\right),\;\;
{\cal G}_{\perp} = m\left(\begin{array}{c} 0 \\ \bm{g}_{\perp} + i(\bm{n}\times\bm{g}_{\perp}) \end{array}\right). 
\end{eqnarray}
We shall now replace the equations of motion by their complex counterpart
\begin{equation} \label{eqnmotc}
\frac{d{\cal W}(t)}{dt} = \hat{\cal M}(\Omega)\!\cdot\!{\cal W}(t) + {\cal G}_{\parallel} + {\cal G}_{\perp}{e}^{i\Omega t}.
\end{equation}
The physical trajectory in phase space is described by the real part of the complex vector ${\cal W}(t)$. Let us introduce a basis of six eigenvectors of $\hat{\cal M}(\Omega)$
\begin{equation}
\hat{\cal M}(\Omega){\cal X}_{k} 
= i \omega_{k}(\Omega){\cal X}_{k},\qquad k = 1,\ldots, 6
\end{equation}
and expand ${\cal W}(t)$ and ${\cal G}(t)$ in this base as follows
\begin{eqnarray}\label{expan}
{\cal W}(t) &=&\sum_{k=1}^{6} \alpha^k(t) \, {e}^{i\omega_k(\Omega)t} \, {\cal X}_k, \\
{\cal G}_\parallel &=& \sum_{k=1}^{6} \gamma^k_{\parallel}\, {\cal X}_{k},\;\;\;
{\cal G}_\perp = \sum_{k=1}^{6} \gamma^k_{\perp}\, {\cal X}_{k}. 
\end{eqnarray}
Owing to a simple block structure of $\hat{\cal M}(\Omega)$, the basis vectors ${\cal X}_{k}$ can be determined by reducing effectively the problem to three dimensions. We use this method in the Appendix B to determine the resonant solution. 

The equation of motion (\ref{eqnmotc}) can be rewritten now as a set of equations for the coefficient functions $\alpha^k(t)$
\begin{equation} \label{alphaeq}
\frac{d{\alpha}^k(t)}{dt} = \gamma^{k}_{\parallel}{e}^{-i \omega_k(\Omega)t} + \gamma^{k}_{\perp}{e}^{i(\Omega-\omega_k(\Omega))t},\;\; k = 1,\ldots,6.
\end{equation}
It is clear now that the mode amplitude $\alpha_k(t)$ will grow linearly in time --- the signature of a resonance --- whenever either one of the two terms on the right hand side becomes time independent. This happens to the first term if one of the frequencies $\omega_k(\Omega)$ vanishes but the corresponding coefficient $\gamma^{k}_{\parallel}$ does not vanish. This case is not interesting, since it means that we are just at the border of the lower instability region and the trap is not holding particles, as discussed in the next section. The second term becomes time independent when the angular velocity of trap rotation $\Omega$ satisfies the resonance condition $\Omega=\omega_k(\Omega)$ and, of course, $\gamma^{k}_{\perp} \neq 0$. This resonance {\em is different} from a resonance in a standard periodically driven oscillator. In the present case the characteristic frequencies of the trap depend on the frequency $\Omega$ of the driving force. Therefore, the position of the resonance has to be determined selfconsistently. A full description of these gravity induced resonances requires the knowledge of the behavior of $\omega_k(\Omega)$'s as functions of $\Omega$. In particular, it is important to know whether a resonance occurs in a region where the system undergoes stable oscillations. This will be discussed in the next Section.

\section{Regions of stability}

\begin{figure}
\centering
\includegraphics[angle=-90,width=0.45\textwidth]{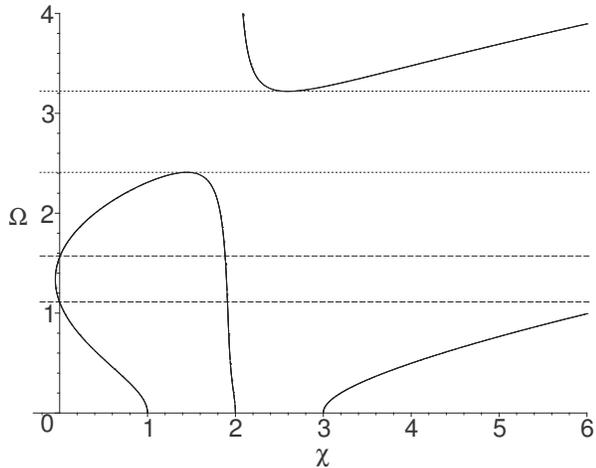}
\caption{The contour lines in this figure represent the zeroes of the characteristic polynomial (\ref{charpoly}) for $V_x=1$, $V_y=2$, $V_z=3$ and $n_x=1/\sqrt{3}$, $n_y=1/\sqrt{3}$, $n_z=1/\sqrt{3}$ plotted as functions of the magnitude of angular velocity $\Omega$ and the square of the characteristic frequency $\chi=\omega^2$. In addition to the lower region of instability (between dashed lines) where one of the roots of (\ref{charpoly}) is negative, there is also an upper region (between dotted lines) where there exists only one real root (determined by the continuation of the line that begins at $\chi=3$, not seen in this figure because it corresponds to a very large value of $\chi$). Since there is no absolute scale of frequencies involved in the analysis of stability regions, we have chosen the lowest trap frequency as a unit.}\label{fig1}
\end{figure}

The stability of motion for a harmonic oscillator is determined by the values of its characteristic frequencies $\omega$ --- the roots of the characteristic polynomial. In the present case, these frequencies are determined by the characteristic equation for the matrix $\hat{\cal M}(\Omega)$
\begin{equation}
\mathrm{Det}\left\{ \hat{\cal M}(\Omega) - i\omega \right\} = 0.
\end{equation}
The characteristic polynomial is tri-quadratic
\begin{equation} \label{charpoly}
Q(\chi)=\chi^3 + A\,\chi^2 + B\,\chi + C,\;\;\;\chi=\omega^2,
\end{equation}
where the coefficients $A$, $B$ i $C$ can be expressed in a rotationally invariant form \cite{cmm}
\begin{eqnarray}
A &\!=\!&-2\Omega^2 - \mathrm{Tr}\{\hat{V}\},\nonumber\\
B &\!=\!&\Omega^4\!+\!\Omega^2(3\bm{n}\!\cdot\!\hat{V}\!\cdot\!\bm{n}\!
-\!\mathrm{Tr}\{\hat{V}\})\!
+\!\frac{\mathrm{Tr}\{\hat{V}\}^2\!-\!\mathrm{Tr}\{\hat{V}^2\}}{2},\nonumber\\
C &\!=\!&{\Omega}^2(\mathrm{Tr}\{\hat{V}\}\!
-\!{\Omega}^2)\bm{n}\!\cdot\!\hat{V}\!\cdot\!\bm{n}\!
-\!\Omega^2\bm{n}\!\cdot\!\hat{V^2}\!\cdot\!\bm{n}\!
-\!\mathrm{Det}\{\hat{V}\}.\qquad
\end{eqnarray}
Stable oscillations take place when all characteristic $\omega$'s are real. This means that all three roots of the polynomial $Q(\chi)$ must be real and positive. Without rotation, when $\Omega=0$, the three roots of $Q(\chi)$ are equal to the eigenvalues of the potential matrix $\hat{V}$. We have then a simple system of three harmonic oscillators vibrating independently along the principal directions of the trap. As $\Omega$ increases, our system will, in general, go through two regions of instability: the lower region when one of the roots of $Q(\chi)$ is negative and the upper region when two roots are complex. We shall exhibit this behavior by plotting the zero contour lines of $Q(\chi)$ in the $\Omega\chi$-plane. We assume that the trap potential and the direction of rotation are fixed and we treat the characteristic polynomial $Q(\chi)$ as a function of $\Omega$ and $\chi$ only. Contour lines representing the zeroes of $Q(\chi)$ in the generic case are shown in Fig.~\ref{fig1}. There is a region of $\Omega$, where only one real root of $Q(\chi)$ exists. However, this region is bounded, so for sufficiently large $\Omega$ the system is always stable.

It has been argued in Ref.~\cite{cmm} that there is always a region of instability when one of the roots of $Q(\chi)$ is negative. The corresponding modes grow exponentially with time. As seen in Fig.~\ref{fig1}, this region of instability is bounded by the two values $\Omega_{1,2}$ at which the curve crosses the vertical axis. These values are given by the zeroes of $C$, treated as a biquadratic expressions in $\Omega$
\begin{equation}\label{zeroes}
\Omega_{1,2} = \sqrt{\frac{b\pm\sqrt{b^2-4ac}}{2a}},
\end{equation}
where $a=\bm{n}\!\cdot\!\hat{V}\!\cdot\!\bm{n},\;
b=\mathrm{Tr}\{\hat{V}\}\bm{n}\!\cdot\!\hat{V}\!\cdot\!\bm{n}
-\bm{n}\!\cdot\!\hat{V^2}\!\cdot\!\bm{n}$, and $c=\mathrm{Det}\{\hat{V}\}$. Since $a, b$, and $c$ are positive and $b^2\geq 4ac$, both values $\Omega_{1}$ and $\Omega_{2}$ are real. A degenerate case is possible, when $\Omega_{1}=\Omega_{2}$ then the region of instability shrinks to zero. In order to determine, when this can happen, we may use the (explicitly non-negative) representation of the discriminant $b^2-4ac$ given in Ref.~\cite{cmm}. Assuming for definitness that $V_x<V_y<V_z$, we find that this happens in two cases: when the trap is not fully anisotropic ($V_x=V_y$ or $V_y=V_z$) or the axis of rotation lies in the $xz$-plane and its azimuthal angle satisfies the condition $\sin^2\theta=(1-V_x/V_y)/(1-V_x/V_z)$. The second possibility has not been noticed in Ref.~\cite{cmm}.

\begin{figure}
\centering
\includegraphics[angle=-90,width=0.45\textwidth]{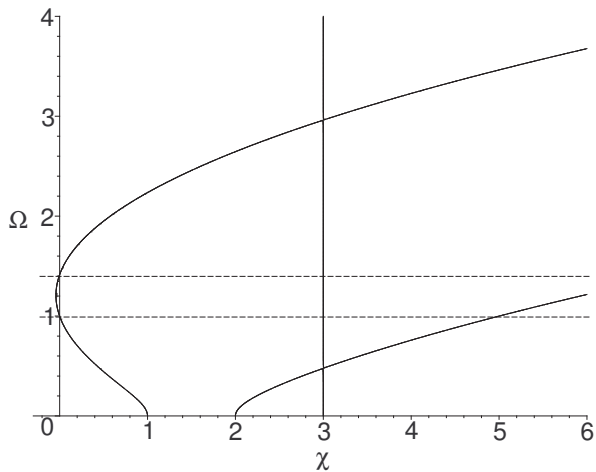}
\caption{This plot is for the same trap as in Fig.~\ref{fig1} but for the rotation around the trap axis $n_x=0$, $n_y=0$, $n_z=1$. In this degenerate case there exists only one region of instability, when one of the roots of (\ref{charpoly}) is negative.} \label{fig2}
\end{figure}

\begin{figure}
\centering
\includegraphics[angle=-90,width=0.45\textwidth]{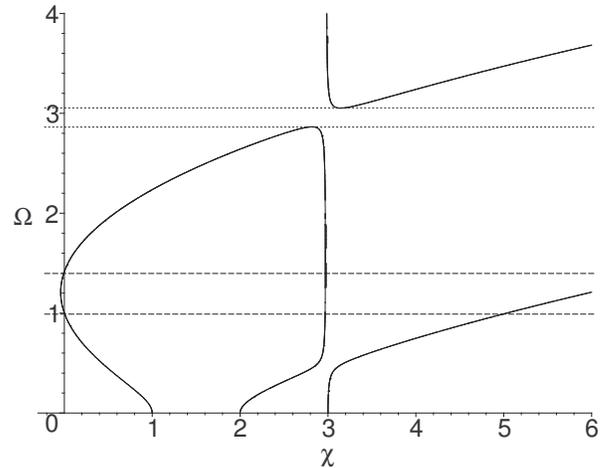}
\caption{This plot shows how the degenerate case (Fig.~\ref{fig2}) merges with the general case (Fig.~\ref{fig1}). Here we have chosen the following values of the parameters: $V_x=1$, $V_y=2$, $V_z=3$ and $n_x=\sin(1/10)$, $n_y=0$, $n_z=\cos(1/10)$.} \label{fig3}
\end{figure}

Graphical representation of the solutions for $n_x=1$, $n_y=n_z=0$ is shown in Fig.~\ref{fig2}. In this plot only one instability region is present, where one of the roots of (\ref{charpoly}) is negative. Owing to the stabilizing effect of the Coriolis force, for fast rotations the system becomes again stable. We can also see that one of the characteristic frequencies remains constant, it does not vary with $\Omega$. It is so, because the rotation does not influence the motion in the direction of the rotation axis. This is the degenerate case described by Whittaker and thoroughly studied in connection with the BEC traps \cite{linn,oktel,gd,ros}. In this case there is only one instability region, where the square of the frequency is negative. When the direction of angular velocity is not parallel to one of the axes of the trap, a second kind of instability appears. In addition to the region of (purely exponential) instability, described before, when one root of $Q(\chi)$ was negative there is also, in general, an additional region of (oscillatory) instability where two roots are complex. Since the coefficients of the characteristic polynomial are real, the square of the second frequency is complex conjugate to the first one. In this case the instability has the form of expanding oscillations. In Fig.~\ref{fig3} we show how the second kind of instability develops when the rotation axis is tilted away a little bit from the $z$ direction.

We will show now, that the second kind of instability does not occur in the most common case, when the rotation vector is along one of the axes of the trap. In particular, it can never happen in a two dimensional trap. Without any loss of generality, we may assume that the direction of rotation is along the $z$ axis. In this degenerate case the characteristic polynomial (\ref{charpoly}) factorizes because the motion in the $z$ direction is not influenced by rotation,
\begin{eqnarray}
\label{factor}
Q(\chi) = (\chi - V_z)\hspace{6cm}\nonumber\\
\times(\chi^2-\chi(2\Omega^2+V_x+V_y)+\Omega^4
-\Omega^2(V_x+V_y)+V_x V_y).\nonumber\\
\end{eqnarray}
The factor quadratic in $\chi$ has real zeros if its discriminant $\Delta$ is positive and this is, indeed, the case since
\begin{eqnarray} \label{delta}
\Delta = 8\Omega^2(V_x+V_y)+(V_x-V_y)^2 \geq 0.
\end{eqnarray}
It follows, however, from the topology of the curves representing the zeros that the second kind of instability always exists if the direction of rotation does not coincide with one of the axes of the harmonic trap. 

\begin{figure}
\centering
\includegraphics[angle=-90,width=0.45\textwidth]{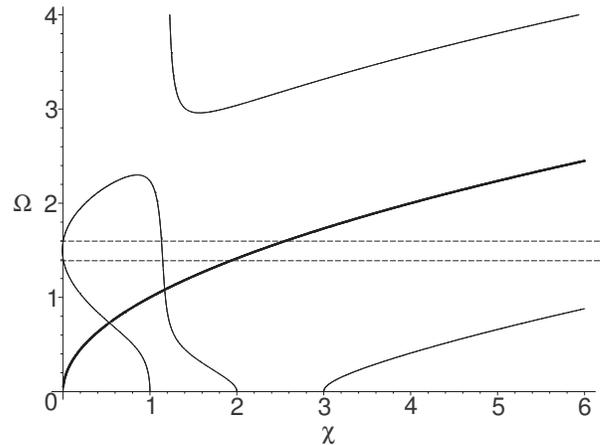}
\caption{$V_x=1$, $V_y=2$, $V_z=3$ $n_x=\sin(\frac{2\pi}{5})$, $n_y=0$, $n_z=\cos(\frac{2\pi}{5})$. Two horizontal lines enclose the lower region of instability. Both resonant frequencies lie in the lower region of stability.} \label{fig4}
\end{figure}

\begin{figure}
\centering
\includegraphics[angle=-90,width=0.45\textwidth]{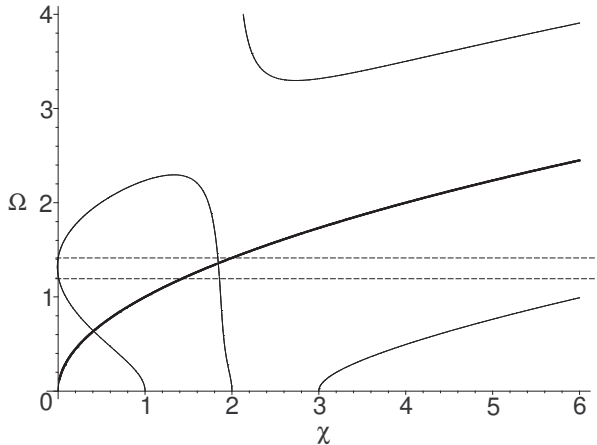}
\caption{$V_x=1$, $V_y=2$, $V_z=3$ $n_x=\sin(\frac{\pi}{4})$, $n_y=0$, $n_z=\cos(\frac{\pi}{4})$. Higher resonant frequency lies in the lower region of instability.} \label{fig5}
\end{figure}
\begin{figure}
\centering
\includegraphics[angle=-90,width=0.45\textwidth]{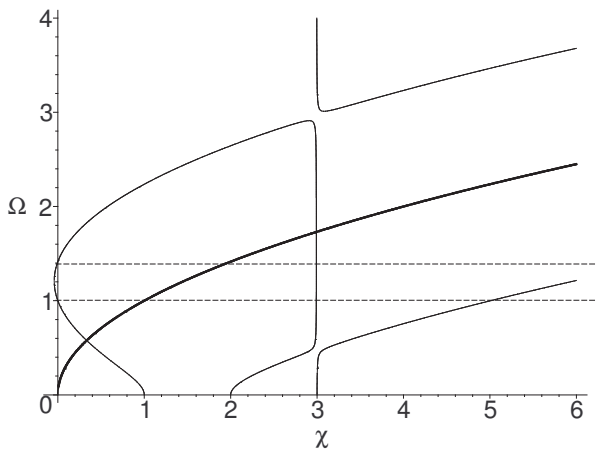}
\caption{$V_x=1$, $V_y=2$, $V_z=3$ $n_x=\sin(\frac{\pi}{60})$, $n_y=0$, $n_z=\cos(\frac{\pi}{60})$. Resonant frequencies lie in two different regions of stability.} \label{fig6}
\end{figure}

\section{Gravity induced resonances}

\begin{figure}
\centering
\includegraphics[angle=-90,width=0.45\textwidth]{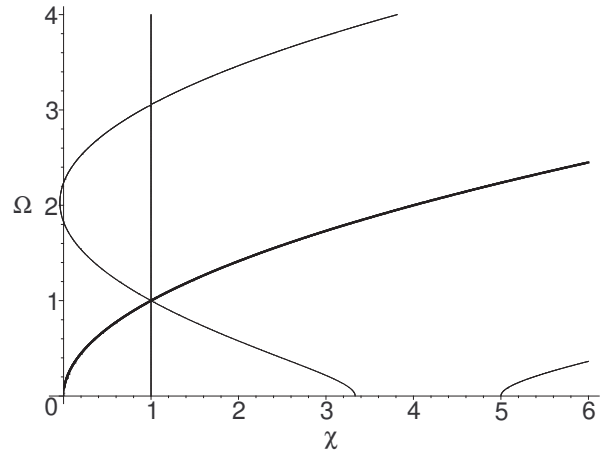}
\caption{In the degenerate case, when two resonant frequencies coincide, their common value is also equal to the lowest frequency of the trap. In this case the trap parameters are: $V_x = 1, V_y = 10/3, V_z = 5$, and $n_x = 1$.}\label{fig7}
\end{figure}

There are, in general, two resonant values of $\Omega$. They occur when $\chi=\Omega^2$. In this case, from Eq.~(\ref{charpoly}) we get a biquadratic equation for the resonant values of $\Omega$:
\begin{equation} \label{reseq}
 D \Omega^4 + E \Omega^2 +F = 0,
\end{equation}
where
\begin{eqnarray}
D &=& -2(\,\mathrm{Tr}\{\hat{V}\}-\bm{n}\!\cdot\!\hat{V}\!\cdot\!\bm{n}\,), \nonumber \\
E &=& \frac{\mathrm{Tr}\{\hat{V}\}^2\!-\!\mathrm{Tr}\{\hat{V}^2\}}{2}
+\mathrm{Tr}\{\hat{V}\}\bm{n}\!\cdot\!\hat{V}\!\cdot\!\bm{n}\! - \bm{n}\!\cdot\!\hat{V^2}\!\cdot\!\bm{n},\nonumber \\
F &=& -\mathrm{Det}\{\hat{V}\}.
\end{eqnarray}
The properties of the two solutions $\Omega_{\pm}^2$ of Eq.~(\ref{reseq}) are described in the Appendix A. We may also analyze the resonance condition graphically by superposing the parabola $\chi=\Omega^2$ (shown as a thick line in Figs.~\ref{fig4}--\ref{fig7}) on the plots of characteristic frequencies. The lower value is more interesting because, as shown in the Appendix A, it always falls into the range of the first region of stable oscillations where we would expect the confinement of particles in the trap. The upper value may fall in the lower region of stability (Fig.~\ref{fig4}), in the lower region of instability (Fig.~\ref{fig5}), or in the higher region of stability (Fig.~\ref{fig6}). A degenerate case is also possible (cf. Fig.~\ref{fig7}) when the two resonance frequencies coincide but this happens only under very special circumstances (see Appendix A). The existence of only two resonant frequencies (we would expect three resonant frequencies for a three dimensional oscillator) clearly shows that we are dealing here with a more complex dynamical system than a simple driven harmonic oscillator.

\begin{figure}
\centering
\includegraphics[width=0.45\textwidth]{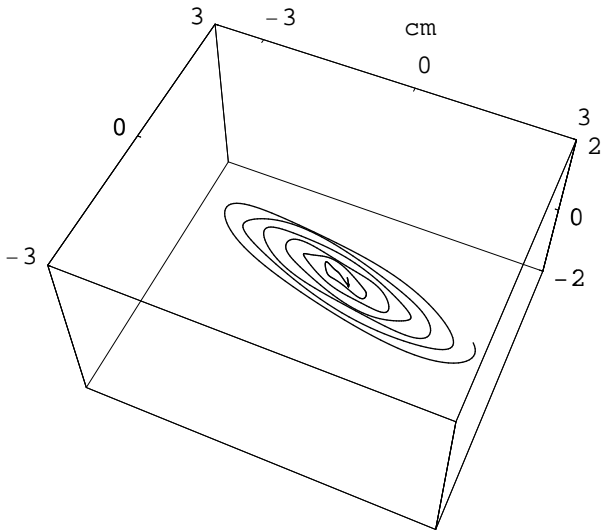}
\caption{The trajectory of a particle in the coordinate frame rotating with the trap. The characteristic frequencies of the trap in the $x$, $y$, and $z$ directions are 10 Hz, 15 Hz, and 20 Hz, respectively. The angular velocity vector has the direction $(1,1,1)$ and its length $\Omega/2\pi = 6.49421$ Hz satisfies the resonance condition, as calculated in the Appendix A. The distances in this plot are measured in centimeters. The direction of the gravitational force ($g=9.81 {\rm m s}^{-2}$) is assumed to coincide at $t=0$ with the $z$-axis of the trap. We can see that the amplitude of oscillations increases linearly with time.}\label{fig8}
\end{figure}

\begin{figure}
\centering
\includegraphics[width=0.45\textwidth]{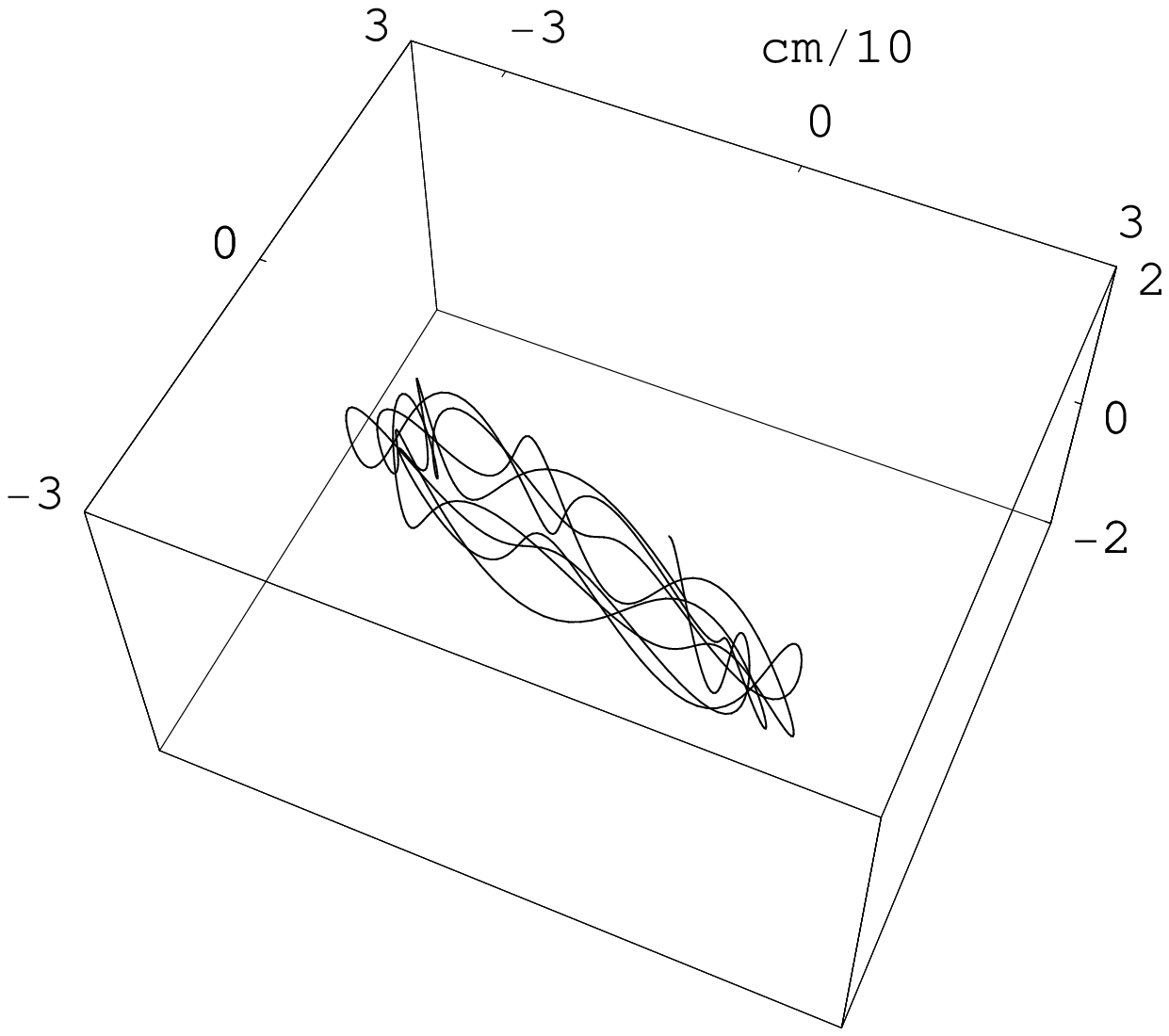}
\caption{The trajectory of a particle under the same conditions as in Fig.~\ref{fig8}, except that the value of the angular velocity  $\Omega/2\pi = 4$ Hz is below the resonance. The scale in this plot is reduced by a factor of 10 as compared to Fig.~\ref{fig8}.}\label{fig9}
\end{figure}

\begin{figure}
\centering
\includegraphics[width=0.45\textwidth]{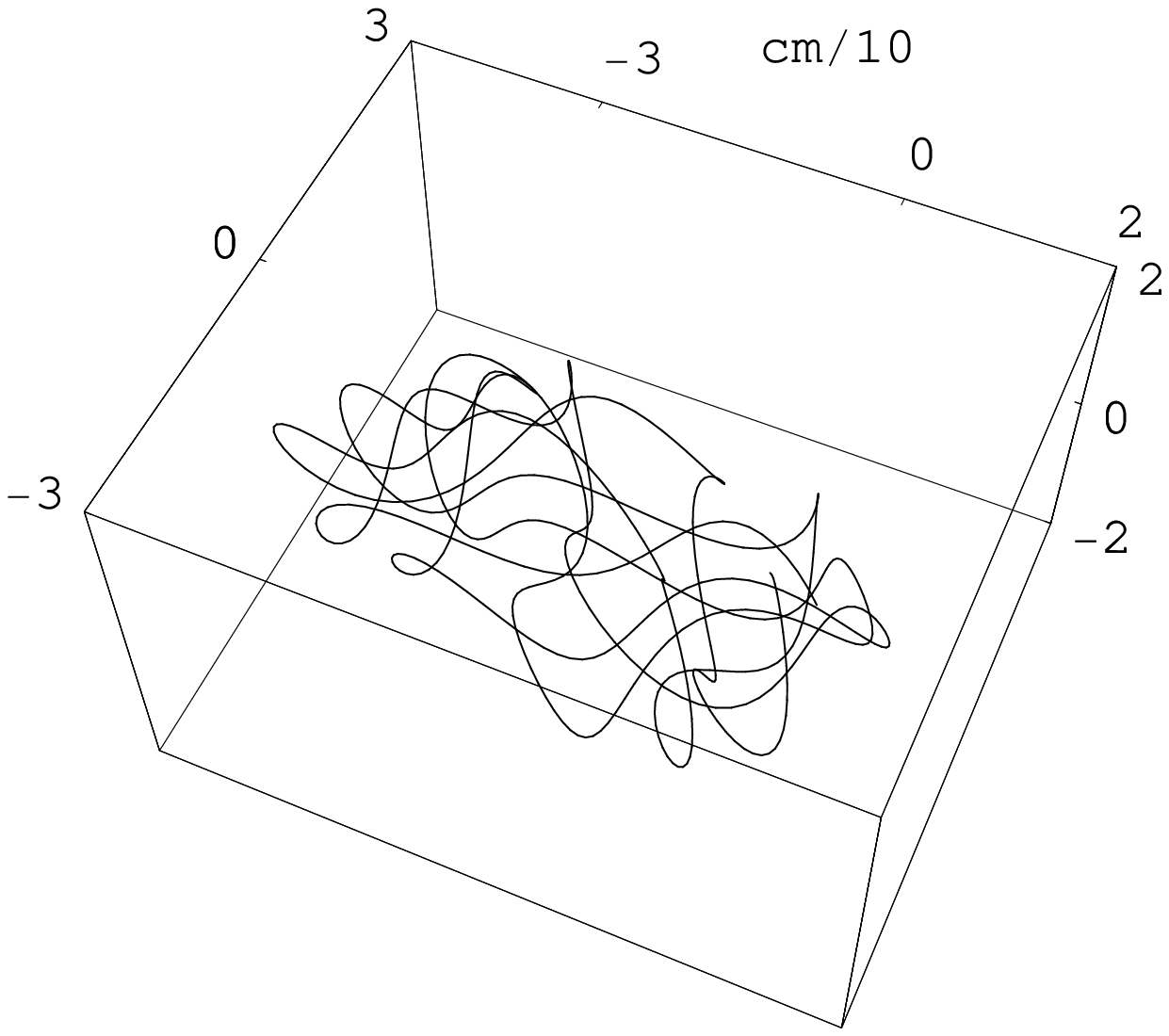}
\caption{The trajectory of a particle under the same conditions as in Fig.~\ref{fig8}, except that the value of the angular velocity $\Omega/2\pi = 9$ Hz is above the resonance. The scale in this plot is reduced by a factor of 10 as compared to Fig.~\ref{fig8}.}\label{fig10}
\end{figure}

\begin{figure}
\centering
\includegraphics[width=0.45\textwidth]{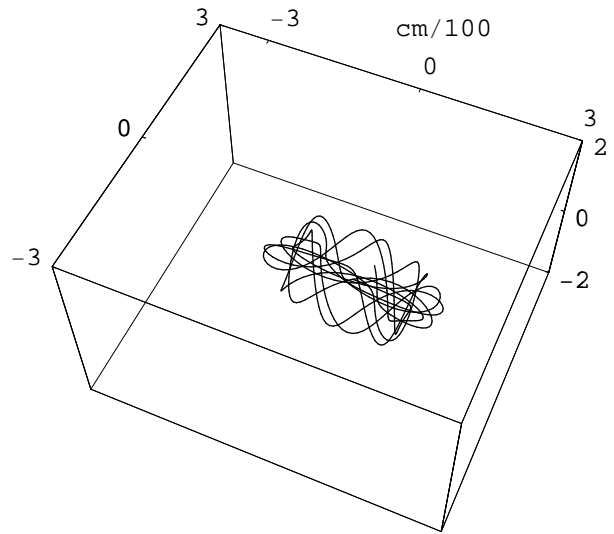}
\caption{The trajectory of a particle under the same conditions as in Fig.~\ref{fig8}, except that the gravitational field has been turned off. The angular velocity $\Omega/2\pi = 6.49421$ Hz has the resonant value. The scale in this plot is reduced by a factor of 100 as compared to Fig.~\ref{fig8}. Clearly, there is no sign of any resonant behavior.}\label{fig11}
\end{figure}

\begin{figure}
\centering
\includegraphics[width=0.45\textwidth]{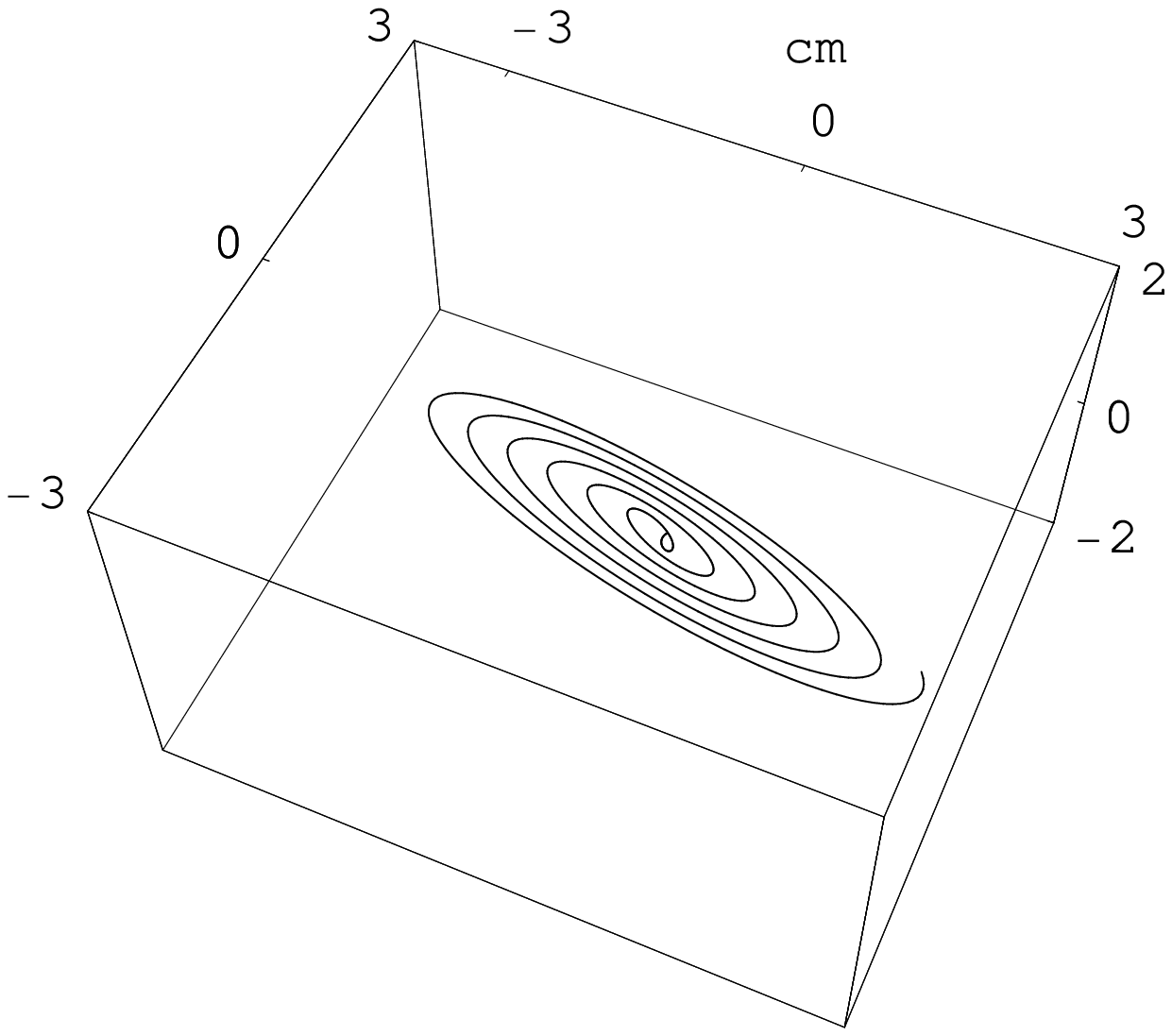}
\caption{The trajectory of a particle described by the analytic formulas given in the Appendix B. The trap parameters and the angular velocity are the same as in Fig.~\ref{fig8}.}\label{fig12}
\end{figure}

The essential difference between the resonant and the nonresonant behavior is illustrated in Figs.~\ref{fig8}-\ref{fig10}. The role of gravity is best seen by comparing Fig.~\ref{fig8} (with gravity) and Fig.~\ref{fig11} (gravity switched off). The calculations were performed in the coordinate system in which the potential matrix $\hat V$ is diagonal and it was assumed that the force of gravity at $t=0$ is directed along the $z$ axis of the trap. All four figures were generated for the same trap and under the same initial conditions: the particle is initially placed in the center of the trap and it is given the initial velocity of 1 cm/s in the $x$ direction. These simple initial conditions result in the excitation of all the modes of the oscillator. However, at resonance the mode growing linearly with $t$ dominates the time evolution. This can be seen by comparing, for the same parameters of the trap and at the resonant frequency $\Omega/2\pi = 6.49421$ Hz, the motion depicted in Fig.~\ref{fig8}, that was calculated numerically, with the motion depicted in Fig.~\ref{fig12}. The second plot was obtained from an analytic solution of the equations of motion (\ref{eqnmotc}), that contains only the resonant part. This analytic solution is derived in the Appendix B. Both plots are essentially the same except for very small differences that are due to the fact that the initial conditions assumed in Fig.~\ref{fig8} require some admixture of the nonresonant solutions while the analytic solution does not contain any nonresonant pieces.

Gravity induced resonances are significant only if they occur in the region of stability. In all regions of instability, where the trajectories exhibit exponential growth, the resonances cannot be detected.

\section{Conclusions}

We have analyzed the stability of motion in an anisotropic, rotating harmonic trap in 3D. We have found that, in general, there are three regions of stability. The second region and the third region merge only when the rotation axis coincides with the one of the trap axes. We have demonstrated the presence of resonances in a rotating harmonic trap subjected to the force of gravity. The nature of these resonances is different from the standard forced harmonic oscillator since the resonant frequencies must be determined selfconsistently. The resonances occur at two rotation rates but they exist only when the rotation axis is tilted away from the vertical direction. However, the rotation axis can still be directed along one of the trap axes. The lower resonance always falls in the region of stable oscillations where as the higher resonance may fall in the lower region of stability, in the first region of instability, or in the higher region of stability. Resonant rotation rates depend solely on the properties of the trap and not on the masses of particles. The resonances cause the escape of a particle, or the center of mass for a collection of interacting particles, from the trap. In view of recent experiments in which the axis of rotation was tilted away from the trap axis \cite{marago,smith}, it should be possible to confirm experimentally the existence of the gravity-induced resonances for Bose-Einstein condensates in rotating traps. In particular, the presence of resonances makes it possible to quickly expel all particles from the trap.

This research was supported by a grant from the Polish Ministry of Scientific Research and Information Technology under the Grant No. PBZ-MIN-008/P03/2003.

\newpage
\appendix
\section{Calculation of resonant angular velocities}

The resonant values of the angular velocity $\Omega$ are the ones that coincide with a characteristic frequency of the trap. Thus, the resonant values of $\Omega$ are obtained by solving the biquadratic equation (\ref{reseq})
\begin{widetext}
\begin{eqnarray}\label{equres}
Q(\Omega^2) \equiv -2\Omega^4({\mathrm{Tr}}\{\hat{V}\} - {\bm n}\!\cdot\!\hat{V}\!\cdot\!{\bm n}) + \Omega^2\left(\frac{{\mathrm{Tr}\{\hat{V}\}^2 - {\mathrm{Tr}}\{\hat{V}}^2\}}{2} + {\mathrm{Tr}}\{\hat{V}\}\,{\bm n}\!\cdot\!\hat{V}\!\cdot\!{\bm n} - {\bm n}\!\cdot\!\hat{V}^2\!\cdot\!{\bm n}\right) - \mathrm{Det}\{\hat{V}\} = 0.
\end{eqnarray}
\end{widetext}
In the coordinate system aligned with the principal axes of the trap, the solutions of this equation have the form
\begin{widetext}
\begin{eqnarray}\label{res}
\Omega^2_\pm = \frac{(1-n_x^2/2)V_y V_z + \dots\pm\sqrt{((1-n_x^2/2)V_y V_z + \dots)^2 - 2V_x V_y V_z((1-n_x^2)V_x + \dots)}}
{2(1-n_x^2)V_x + \dots},
\end{eqnarray}
\end{widetext}
where $V_x, V_y, V_z$ denote the diagonal elements of $\hat{V}$ and the dots stand everywhere for two additional terms obtained by the cyclic substitutions $x\to y\to z\to x$. These solutions are always real since the discriminant $\Delta$ --- the expression under the square root --- is never negative. This is seen from the following representation of $\Delta$ as a sum of two nonnegative terms (for definitness, we have assumed here that $V_x<V_y<V_z$)
\begin{widetext}
\begin{eqnarray}
\Delta &=& \left((1-n_x^2/2)V_y V_z-(1+n_y^2/2)V_x V_z
-(1+n_z^2/2)V_x V_y\right)^{2}\nonumber\\
&+&4V_x\left(n_z^2(V_z-V_x)(V_y V_z+V_y^2/2)
+n_y^2(V_y-V_x)(V_y V_z+V_z^2/2)\right).
\end{eqnarray}
\end{widetext}
This representation of $\Delta$ can also be used to find the necessary condition for the two resonant values to merge into one. The second term vanishes (for a nondegenerate trap) only if $n_x=\pm 1$ and the first term then vanishes when the trap frequencies satisfy the condition (cf. Fig.~\ref{fig7}.
\begin{eqnarray}
\frac{1}{2 V_x} = \frac{1}{V_y}+\frac{1}{V_z}.
\end{eqnarray}
The lower value $\Omega_-$ of the resonant angular velocity never exceeds the lowest critical angular velocity at which the system becomes unstable. To prove this, we may compare the expression (\ref{equres}) with the following expression that determines the critical value \cite{cmm}
\begin{widetext}
\begin{eqnarray}\label{crit}
 Q(0) \equiv - {\Omega}^4\,{\bm n}\!\cdot\!\hat{V}\!\cdot\!{\bm n} + \Omega^2({\mathrm{Tr}}\{\hat{V}\}\,{\bm n}\!\cdot\!\hat{V}\!\cdot\!{\bm n} - {\bm n}\!\cdot\!\hat{V}^2\!\cdot\!{\bm n}) - \mathrm{Det}\{\hat{V}\} = 0.
\end{eqnarray}
\end{widetext}
Both these expressions may be represented by inverted parabolas that cross the $y$-axis at the same negative value $- \mathrm{Det}\{\hat{V}\}$. Since the derivative of (\ref{equres}) with respect to $\Omega^2$ at the crossing point is larger than the derivative of (\ref{crit}), the parabola (\ref{equres}) is steeper at the crossing point. Therefore, it must cross the $x$-axis at a lower value of $\Omega^2$ and the lower resonant value lies below the boundary of the stability region.

\section{Analytic resonant solution}

A resonant solution is the one that has in the mode expansion (\ref{expan}) only the term oscillating with the resonant frequency $\Omega$. The amplitude of these oscillations is growing linearly in time. In order to find an explicit form of this solution, we substitute into the equations of motion (\ref{eqnmotc}) the following Ansatz
\begin{eqnarray}
{\cal W}_r = ({\cal A}t + {\cal B})e^{i\Omega t} + {\cal C},
\end{eqnarray}
where $\cal A$, $\cal B$, and $\cal C$ are six-dimensional vectors. Upon substituting this form into the equations of motion and comparing the terms that have the same time dependence, we obtain the following equations for the three unknown vectors
\begin{eqnarray}
{\hat\cal M}(\Omega)\!\cdot\!{\cal A} = i\Omega{\cal A},\\
{\hat\cal M}(\Omega)\!\cdot\!{\cal B} = i\Omega{\cal B} + {\cal A} - {\cal G}_\perp,\\
{\hat\cal M}(\Omega)\!\cdot\!{\cal C} = -{\cal G}_\parallel.
\end{eqnarray}
The first equation says that the vector ${\cal A}$ is directed along the resonant eigenmode. From the second equation we can determine the length of this vector. The vector ${\cal B}$ can be determined from the second equation only up to a component along ${\cal A}$. This is so, because by changing the origin of the time scale, we may always add such a component to ${\cal B}$. Finally, the vector ${\cal C}$ is determined from the third equation. It is now a matter of pure algebra to find these three vectors. Six-dimensional problems are quite cumbersome but we may replace them here by their three-dimensional counterparts, owing to a simple block structure of the matrix ${\hat\cal M}(\Omega)$. In addition, let us note that we need only the upper three components of the solution ${\cal W}_r(t)$ to determine the trajectory. We shall call them $\bm a$, $\bm b$, and $\bm c$, respectively. Thus the physical trajectory for the resonant solution will have the form ${\bm r}(t) = \Re(({\bm a}t + {\bm b})e^{i\Omega t} + {\bm c})$. Eliminating the lower three components, we obtain the following set of three-dimensional equations
\begin{subequations}\label{leq}
\begin{eqnarray}
{\hat N}(\Omega)\!\cdot\!{\bm a} = 0,\label{leqa}\\
{\hat N}(\Omega)\!\cdot\!{\bm b} = i\Omega{\bm b} + {\bm a} - {\bm h},\label{leqb}\\
{\hat N}(0)\!\cdot\!{\bm c} = -{\bm g}_\parallel,\label{leqc}
\end{eqnarray}
\end{subequations}
where ${\hat N}(\omega) = \omega^2  - 2i\omega\hat{\Omega} - \hat{\Omega}^2 - {\hat V}$ and ${\bm h} = \bm{g}_{\perp} + i(\bm{n}\times\bm{g}_{\perp})$.

\begin{widetext}
\begin{eqnarray}\label{matn}
{\hat N}(\omega) = 
\begin{pmatrix}
\omega^2+\Omega^2(1-n_x^2)-V_x & 
-\Omega^2 n_x n_y - 2i\omega\Omega n_z) &
 -\Omega^2 n_x n_z + 2i\omega\Omega n_y)\\
 -\Omega^2 n_x n_y + 2i\omega\Omega n_z) & \omega^2+\Omega^2(1-n_y^2)-V_y & 
 -\Omega^2 n_y n_z - 2i\omega\Omega n_x)\\
 -\Omega^2 n_x n_z - 2i\omega\Omega n_y) & 
 -\Omega^2 n_y n_z + 2i\omega\Omega n_x) & \omega^2+\Omega^2(1-n_z^2)-V_z \\
\end{pmatrix}.
\end{eqnarray}
\end{widetext}
The best way to find solutions of Eqs.~(\ref{leq}) is to expand the vectors $\bm{a}$ and $\bm{b}$ in the basis of the eigenvectors of the matrix $\hat N(\Omega)$
\begin{subequations}
\begin{eqnarray}\label{ab}
\bm{a} &=& \alpha_0\bm{e}_0+\alpha_+\bm{e}_+ + \alpha_-\bm{e}_-,\\
\bm{b} &=& \beta_0\bm{e}_0+\beta_+\bm{e}_+ + \beta_-\bm{e}_-.
\end{eqnarray}
\end{subequations}
Upon substituting these formulas into Eq.~(\ref{leqa}) one finds that $\alpha_+$ and $\alpha_-$ vanish, $\beta_0$ is arbitrary, and the remaining three coefficients can be calculated from Eq.~(\ref{leqb}). Everywhere in this Appendix $\Omega$ stands for one of the two resonant frequencies given by the expression (\ref{res}). The characteristic polynomial of the matrix ${\hat N}(\Omega)$ is
\begin{widetext}
\begin{eqnarray}
\lambda^3 + \lambda^2(V_x+V_y+V_z - 5\Omega^2)
 + 4\lambda(\Omega^4-3\Omega^2(V_x+V_y+V_z)+V_yV_z+V_xV_z+V_xV_y).
\end{eqnarray}
\end{widetext}
One of the eigenvalues vanishes and the remaining two eigenvalues $\lambda_\pm$ have the form
\begin{widetext}
\begin{eqnarray}
\lambda_\pm = \frac{5\Omega^2-V_x-V_y-V_z}{2}
\pm\frac{\sqrt{9\Omega^4+2(V_x(1+2n_x^2)+\dots)+V_x^2-2V_yV_z+\dots}}{2},
\end{eqnarray}
\end{widetext}
where the dots have the same meaning as in the Appendix A. Knowing the eigenvalues, we may write down explicit expressions for the projectors ${\hat P_0}, {\hat P_+}$, and ${\hat P_+}$ corresponding to the eigenvectors of ${\hat N}(\Omega)$,
\begin{subequations}
\begin{eqnarray}
{\hat P_0} &=& \frac{({\hat N}(\Omega)-\lambda_+)({\hat N}(\Omega)-\lambda_-)}{\lambda_+\lambda_-},\\
{\hat P_+} &=& \frac{{\hat N}(\Omega)({\hat N}(\Omega)-\lambda_-)}{\lambda_+(\lambda_+-\lambda_-)},\\
{\hat P_-} &=& \frac{{\hat N}(\Omega)({\hat N}(\Omega)-\lambda_+)}{\lambda_-(\lambda_--\lambda_+)}.
\end{eqnarray}
\end{subequations}
The eigenvectors ${\bm e}_0$ and ${\bm e}_\pm$ may be obtained by acting with the projectors on any generic vector (not an eigenvector of ${\hat M}(\Omega)$). We shall choose ${\bm h} = {\bm g}_{\perp} +  i({\bm n}\times{\bm g}_{\perp})$ as this vector because ${\bm h}$ appears already in Eq.~(\ref{leqc}). Thus, the (unnormalized) eigenvectors, needed in the expansions (\ref{ab}), can be written in the form
\begin{subequations}
\begin{eqnarray}
{\bm e}_0 &=& {\hat P_0}\!\cdot\!{\bm h},\\
{\bm e}_+ &=& {\hat P_+}\!\cdot\!{\bm h},\\
{\bm e}_- &=& {\hat P_-}\!\cdot\!{\bm h}.
\end{eqnarray}
\end{subequations}
Taking into account the orthogonality of the eigenvectors, we may find the coefficients $\alpha_0$ and $\beta_\pm$ that lead to the following explicit expression for the vectors ${\bm a},{\bm b}$
\begin{subequations}
\begin{eqnarray}
{\bm a} &=& \!\frac{{\bm e}_0^\dagger\!\cdot\!{\bm e}_0}{2{\bm e}_0^\dagger\!\cdot\!(i\Omega+{\hat\Omega})\!\cdot\!{\bm e}_0}\;{\bm e}_0,\\
{\bm b} &=& \!\left(\frac{i({\bm e}_0^\dagger\!\cdot\!{\bm e}_0)\;({\bm e}_+^\dagger\!\cdot\!{\hat\Omega}\!\cdot\!{\bm e}_0)}{({\bm e}_0^\dagger\!\cdot\!(i\Omega+{\hat\Omega})\!\cdot\!{\bm e}_0)\;({\bm e}_+^\dagger\!\cdot\!{\bm e}_+)} - 1\!\right)\!\frac{{\bm e}_+}{\lambda_+}\nonumber\\
&+& \!\left(\frac{i({\bm e}_0^\dagger\!\cdot{\bm e}_0)\;({\bm e}_-^\dagger\!\cdot\!{\hat\Omega}\!\cdot\!{\bm e}_0)}{({\bm e}_0^\dagger\!\cdot\!(i\Omega+{\hat\Omega})\!\cdot\!{\bm e}_0)\;({\bm e}_-^\dagger\!\cdot\!{\bm e}_-)} - 1\!\right)\!\frac{{\bm e}_-}{\lambda_-}.
\end{eqnarray}
\end{subequations}
Finally, the vector ${\bm c}$ is found by solving the equations ${\hat M}(0)\!\cdot\!{\bm c}=-{\bm n}({\bm n}\!\cdot\!{\bm g})$ and it has the following components
\begin{widetext}
\begin{eqnarray}
{\bm c} = \frac{({\bm n}\!\cdot\!{\bm g})\,[n_x(V_y-\Omega^2)(V_z-\Omega^2),\,n_y(V_z-\Omega^2)(V_x-\Omega^2),\,
n_z(V_x-\Omega^2)(V_y-\Omega^2)]}
{V_xV_yV_z-\Omega^2(n_x^2V_x(V_y+V_z)+n_y^2V_y(V_z+V_x)+n_z^2V_z(V_x+V_y))+
\Omega^4(n_x^2V_x+n_y^2V_y+n_z^2V_z)}.
\end{eqnarray}
\end{widetext}

\end{document}